\documentclass[]{spie}  

\usepackage{graphicx}

\def\etal{{\it et al}.}

\def\degree{{${}^\circ$}}

\begin{document}

\thispagestyle{empty}
\renewcommand{\thefootnote}{\fnsymbol{footnote}}

\begin{flushright}
{\small
SLAC--PUB--12508\\
May, 2007\\}
\end{flushright}

\vspace{.8cm}

\begin{center}
{\bf\large   
Studies of EGRET sources with a novel image restoration technique
\footnote{This work was supported in part by the U.S. Department of Energy under Grant DE-AC02-76SF00515.}}

\vspace{1cm}

Hiroyasu~Tajima${}^*$, Stefano~Finazzi${}^\dagger$, Johann~Cohen-Tanugi${}^*$, James~Chiang${}^{*,**}$ and Tuneyoshi~Kamae${}^*$

\medskip

{${}^*$Stanford Linear Accelerator Center, Stanford, CA 94309-4349, USA\\
${}^\dagger$Scuola Normale Superiore, I-56100 Pisa, Italy\\
${}^{**}$CRESST, University of Maryland, Baltimore County, 1000 Hilltop Circle, Baltimore, MD 21250, USA}

\end{center}

\vfill

\begin{center}
{\bf\large   
Abstract }
\end{center}

\begin{quote}
We have developed an image restoration technique based on the Richardson-Lucy algorithm optimized for GLAST-LAT image analysis. Our algorithm is original since it utilizes the PSF (point spread function) that is calculated for each event. This is critical for EGRET and GLAST-LAT image analysis since the PSF depends on the energy and angle of incident gamma-rays and varies by more than one order of magnitude. EGRET and GLAST-LAT image analysis also faces Poisson noise due to low photon statistics. Our technique incorporates wavelet filtering to minimize noise effects.
We present studies of EGRET sources using this novel image restoration technique for possible identification of extended gamma-ray sources.
\end{quote}

\bigskip


\vfill

\begin{center} 
{\it Presented at} 
{\it First GLAST Symposium, Stanford University, Stanford, CA, USA, February 5--8, 2007} \\



\end{center}

\clearpage
\pagestyle{plain}

\section{ INTRODUCTION}
The EGRET (Energetic Gamma Ray Experiment Telescope) instrument onboard the CGRO (Compton Gamma Ray Observatory) detected 271 gamma-ray sources in the energy band 0.1--10~GeV\cite{Hartman99}.
However, 170 of these sources do not have clear counterparts at other wavelength and many of them still remain unidentified.
Most of unidentified EGRET sources are located at low Galactic latitude and are considered to be of Galactic origin.
Although some of them can be extended sources such as supernova remnants, it is difficult to identify extended sources due both to poor PSF (point-spread function) of the EGRET and to the presence of the Galactic diffuse gamma-ray background.
An earlier attempt to improve the EGRET image using Richardson-Lucy deconvolution technique\cite{RL} with wavelet denoising
by Charalabides \etal\cite{Charalabides} suffered from a rapidly varying and energy dependent PSF.
We have developed a novel image restoration technique that incorporates the PSF calculated for each event as described in the following section.
This technique will be critical for analysis of extended sources in the LAT (Large Area Telescope) instrument of the GLAST (Gamma-ray Large Area Telescope) since its PSF varies by two order of magnitude over the wide energy band, 0.1--300~GeV.

\section{ IMAGE RESTORATION TECHNIQUE}
The observed image $\tilde{\phi}(x)$ is a convolution of the true image $\psi(\xi)$ and the instrument response function $P(x|\xi)$, where $P(x|\xi)$ is the probability that the photon is observed at $x$ when the true position is $\xi$.
Based on Bayesian approach, the original image $\psi(\xi)$ can be obtained iteratively:
\begin{equation}
\psi^{r+1}(\xi) = \frac{1}{N}\psi^r(\xi)\sum_{k=1}^{N}\frac{P_k(x_k|\xi)}{\int P_k(x_k|\zeta)\psi^r(\zeta)d\zeta},\hspace{1cm}
\left[\psi^{r+1}(\xi) = \int\tilde{\phi}(x)\frac{\psi^r(\xi)P(x|\xi)}{\int P(x|\zeta)\psi^r(\zeta)d\zeta}dx\right],
\end{equation}
where $x_k$ is the observed position of the $k$th event and $P_k(x_k|\xi)$ is the probability to observe the event at $x_k$ when the true position is $\xi$.
(The first formula is generalized for the case where the PSF is assigned for each event, while the second formula in the square brackets is the original version shown as a reference.)
It is proven mathematically that $\psi^{r+1}(\xi)$ yields larger likelihood than $\psi^{r}(\xi)$ at each iteration.
However, it suffers from the amplification of the noise (mostly Poisson noise due to poor statistics in the EGRET analysis) as the number of iterations is increased.
Efforts have been made to define the objective criteria to stop the iteration before such effects become prominent\cite{Lucy94}.
Wavelet denoising can be used to suppress the noise in the residual between the observed and expected image in each iteration\cite{Starck94}.
The residual, $\rho^r(x)$, is defined as $\rho^r(x) \equiv \tilde{\phi}(x) - \phi^r(x)$, where $\phi^r(x)$ is the convolved image defined as:
\begin{equation}
\phi^r(x) \equiv \int \Pi(x|\zeta)\psi^r(\zeta)d\zeta,\ \ \ 
\Pi(x|\zeta) = \frac{\sum_{k=1}^{N}P_k(x|\zeta)P_k(x_k|x)}{\sum_{k=1}^{N} P_k(x_k|x)},\hspace{1cm}
\left[\phi^r(x) \equiv \int P(x|\zeta)\psi^r(\zeta)d\zeta\right].
\end{equation}
Here $\Pi(x|\zeta)$ is the weighted average of the instrument response at a given position $x$.
The residual is decomposed to $J$ wavelet scales as $\rho^r(x) = c_J(x) + \sum_{j=1}^{J}w_j(x)$,
where $c_J(x)$ is the last smoothed image and $w$ denotes each wavelet scale.
The residual can be filtered based on its significance as:
\begin{equation}
\overline{\rho}^r(x) = c_J(x) + \sum_{j=1}^{J}M_j(x)w_j(x),
\end{equation}
where $M_j(x)$ is 0 if the $w_j(x)$ is consistent with noise and 1 otherwise.
We currently use the 99\% Poisson probability as the consistency threshold.
The filtered residual is incorporated in the iteration process using the filtered image, $\overline{\phi}^r(x) = \overline{\rho}^r(x) + \phi^r(x)$:
\begin{equation}
\psi^{r+1}(\xi) = \frac{1}{N}\psi^r(\xi)\sum_{k=1}^{N}\frac{P_k(x_k|\xi)\overline{\phi}^r(x_k)/\tilde{\phi}(x_k)}{\int P_k(x_k|\zeta)\psi^r(\zeta)d\zeta},\hspace{1cm}
\left[\psi^{r+1}(\xi) = \int\overline{\phi}^r(x)\frac{\psi^r(\xi)P(x|\xi)}{\int P(x|\zeta)\psi^r(\zeta)d\zeta}dx\right],
\end{equation}
The generalized formulas described here are completely equivalent to the non-generalized versions if $P_k(x|\xi)$ is common for all events.

When we have known point sources, the probability density function, $\psi^{r}(\xi)$, can be decomposed into two components: $\psi^r(\xi) = \psi_{\mathrm PS}^r(\xi) + \psi_{\mathrm DF}^r(\xi)$, where $\psi_{\mathrm PS}^r(\xi)$ represents a point source component (mostly zero apart from $\delta$-functions at the positions of the point sources) while $\psi_{\mathrm DF}^r(\xi)$ represents a diffuse component\cite{Hook94}.
Wavelet filtering is applied only to the diffuse component.
This dual channel method enables us to incorporate sharp point sources while keeping the smooth diffuse component.
This effectively works as point source subtraction without negative values in $\psi_{\mathrm DF}^{r}(\xi)$ (straight point source subtraction often causes negative values due to noise).

\section{ APPLICATION TO EGRET IMAGE ANALYSIS}
In this section, we demonstrate our image restoration technique by analyzing three fields observed by the EGRET.
The PSF is calculated for each event as a convolution of PSFs in different energy bins, according to the energy dispersion of the given energy measurement.
Only photons with energy greater than 0.5~GeV are used in this analysis in order to minimize the processing time without sacrificing the image quality.
Lower energy photons do not contribute to improving the quality of the restored image.
(It does not degrade the image quality either since such photons are weighted appropriately via large PSF).
The results presented in this paper are still preliminary and should not be used for any quantitative analysis.

\subsection{ Large Magellanic Cloud}
\begin{figure}[hb]
\centering
  \includegraphics[height=3.9cm]{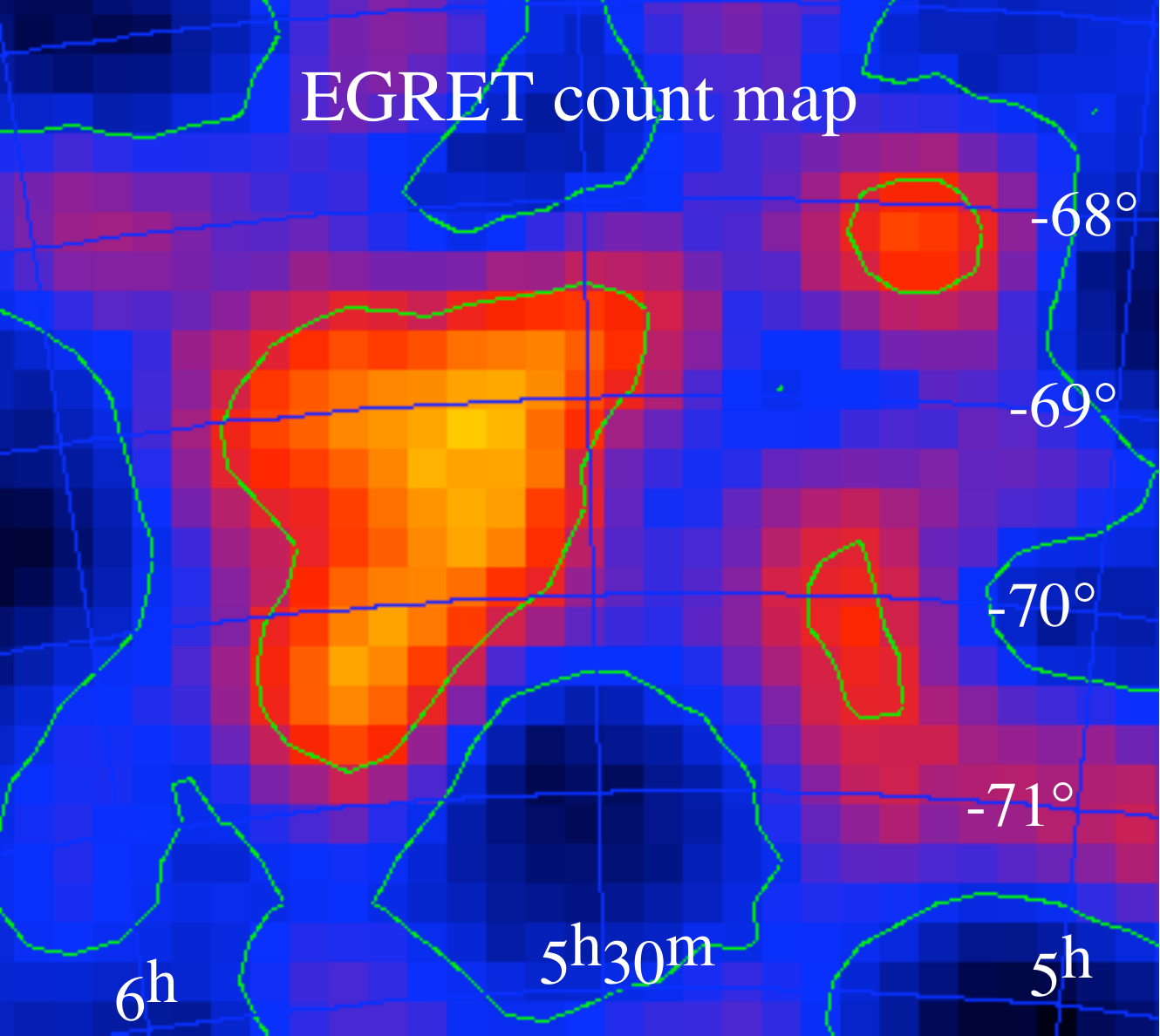}\hspace{0.3cm}
  \includegraphics[height=3.9cm]{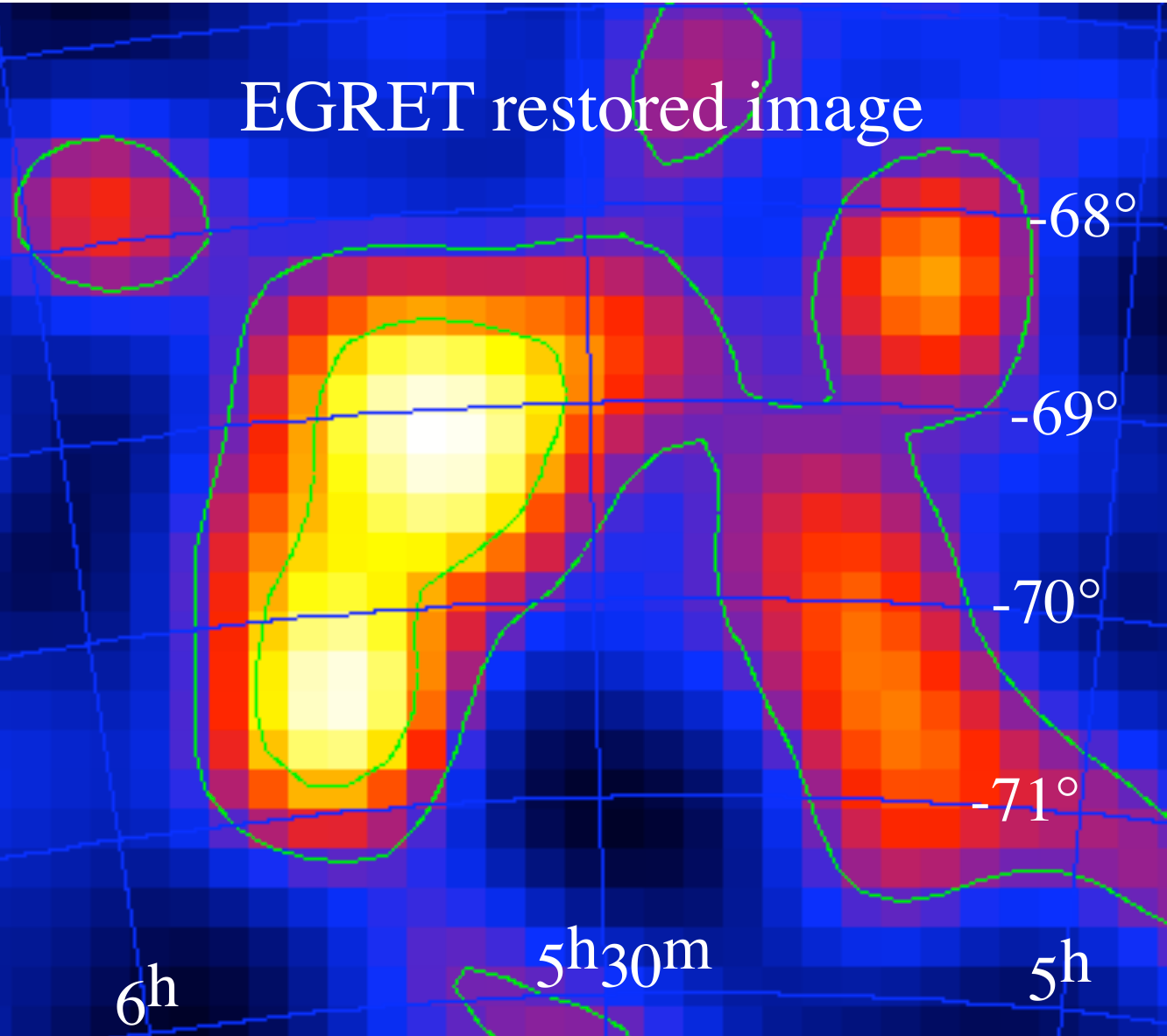}\hspace{0.3cm}
  \includegraphics[height=3.9cm]{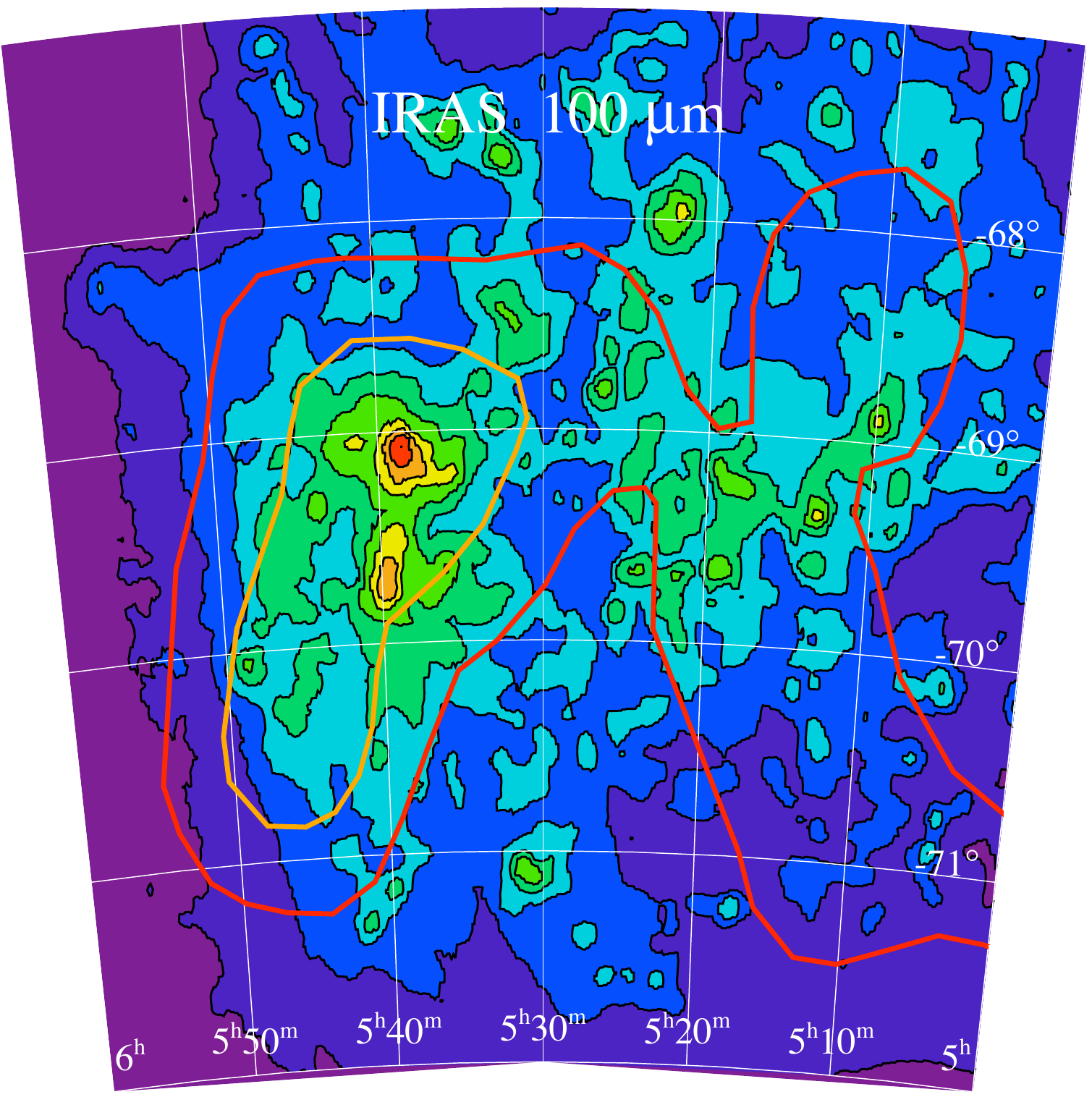}
  \caption{\label{fig:LMC}Smoothed count map (left), restored image (middle) of the LMC by EGRET, and the IRAS LMC image overlaid with contour of the restored EGRET image (right).}
\end{figure}

The LMC (Large Magellanic Cloud) is one of few extended sources resolved by EGRET. 
Figure~\ref{fig:LMC} shows a smoothed photon count map of the LMC in the left panel, a restored image of the LMC in the middle panel, and the IRAS LMC image overlaid with the contour of the restored EGRET image in the right panel.
The restored EGRET image is apparently sharper with better contrast than the original count map while avoiding enhancement of spurious structures (sometimes suppressing them) in the background (The color scale is common for both images).
The bright spot of the restored image is well correlated with a dense region indicated by the IRAS image, which traces dusty interstellar clouds and associated star-forming regions.

\subsection{ 3EG J1234-1318}
\begin{figure}[thb]
\centering
  \includegraphics[height=3.9cm]{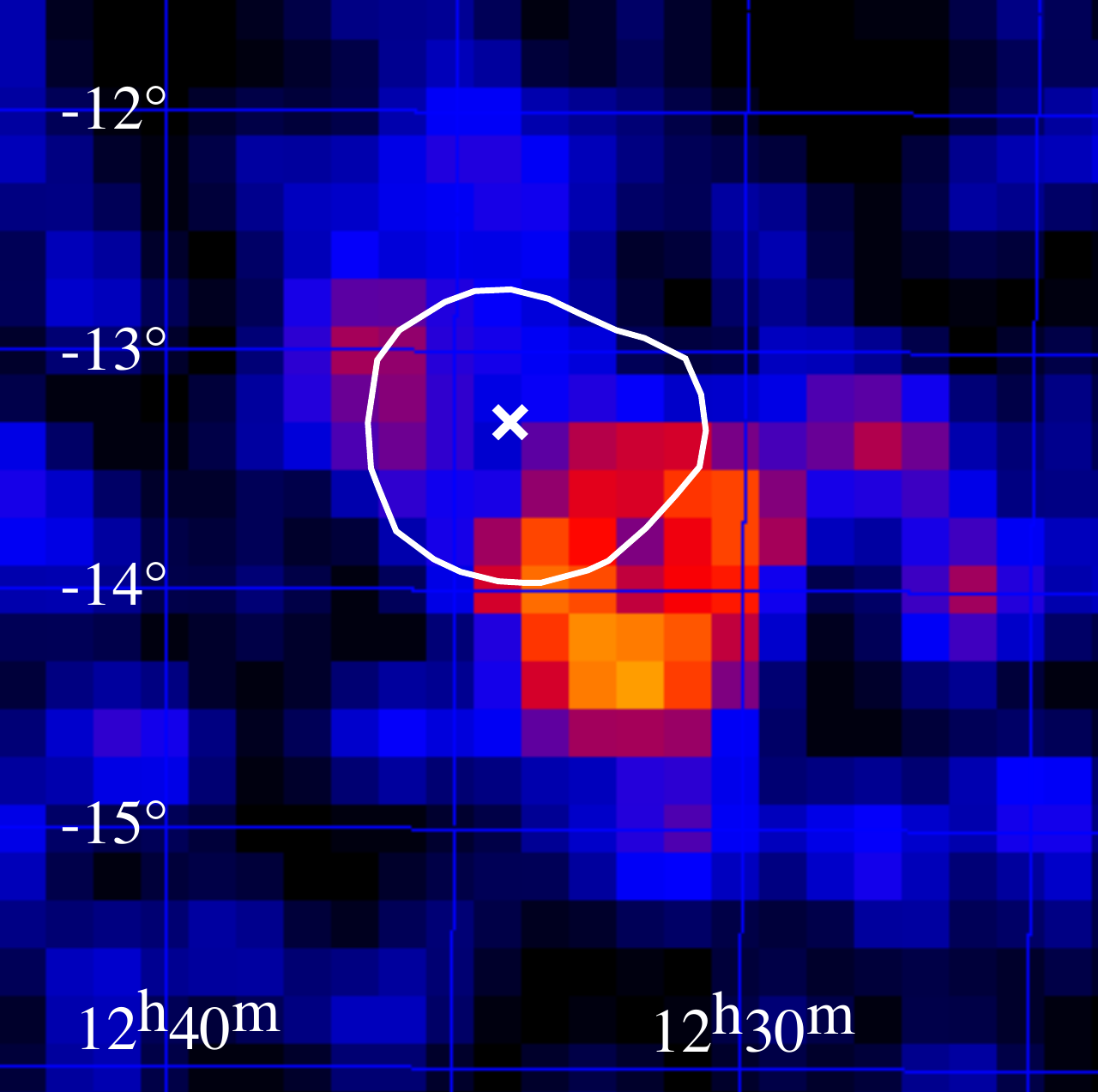}\hspace{0.3cm}
  \includegraphics[height=3.9cm]{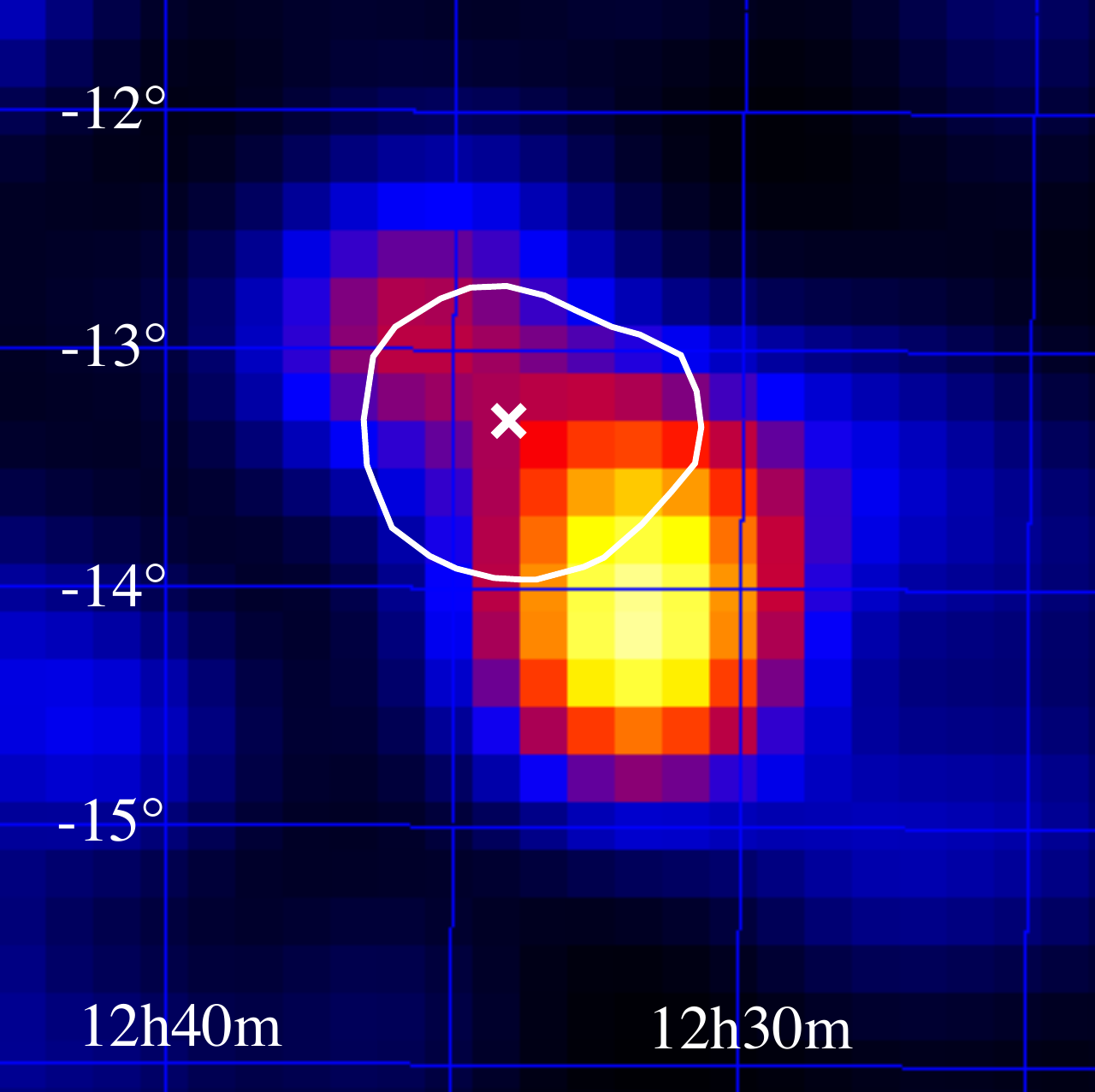}
  \caption{\label{fig:EG1234-1318}Smoothed count map (left) and restored image (right) of the 3EG J1234-1318 by EGRET.}
\end{figure}

We also analyzed several unidentified EGRET sources at high Galactic latitude where the background diffuse component is very low.
We found one source that is inconsistent with one point source as shown in Figure~\ref{fig:EG1234-1318}.
The left panel shows the smoothed count map of the 3EG J1234-1318 while the right panel shows the restored image.
A white cross and a contour in each image indicate the source location and its 90\% confidence level contour in the original 3EG catalog (The 3EG catalog is produced using photons with $E>0.1$~GeV).
The restored image has much better contrast and clearly shows elongated structure indicating the possibility of two point sources with a distance of $\sim$1.5\degree or an extended source.
Since the Galactic latitude of this source is $\sim$50\degree\, it is very likely to be two extra-Galactic point sources.
If it is the case, new point source position would be preliminarily (RA, DEC)=($187.9$\degree, $-14.1$\degree) for the brighter one and ($188.9$\degree, $-13.0$\degree) for the dimmer one instead of the original position, ($188.5$\degree, $-13.3$\degree).
Another possibility is that this is due to a nearby molecular cloud interacting with cosmic-rays, which could be extremely useful to obtain cosmic-ray flux (spectrum and intensity) in that region.
This result demonstrates that our image restoration technique can be quite effective in identification of extended sources and source confusions.

\subsection{ Galactic Plane}
\begin{figure}[hb]
\centering
  \includegraphics[height=3.6cm]{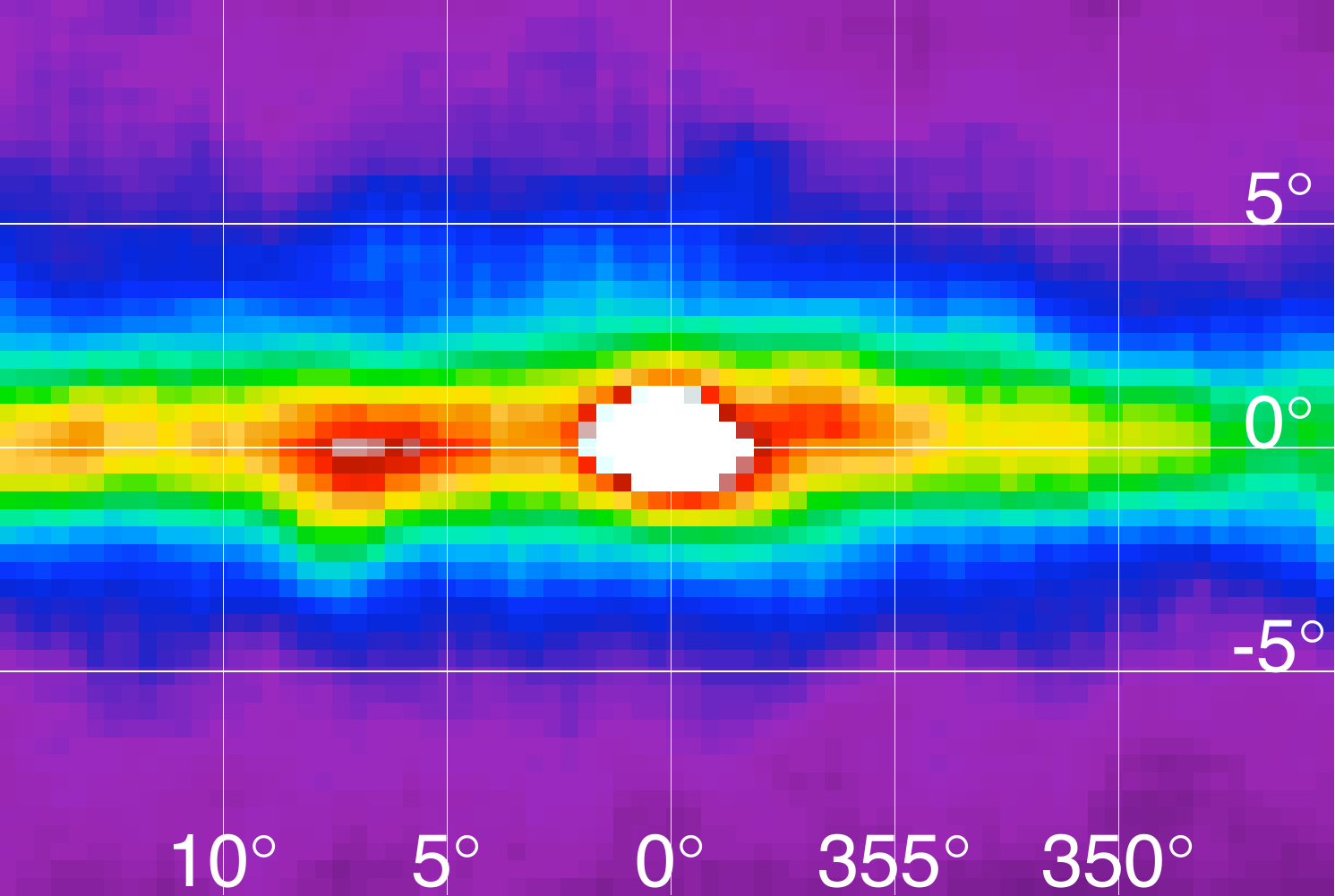}
  \includegraphics[height=3.6cm]{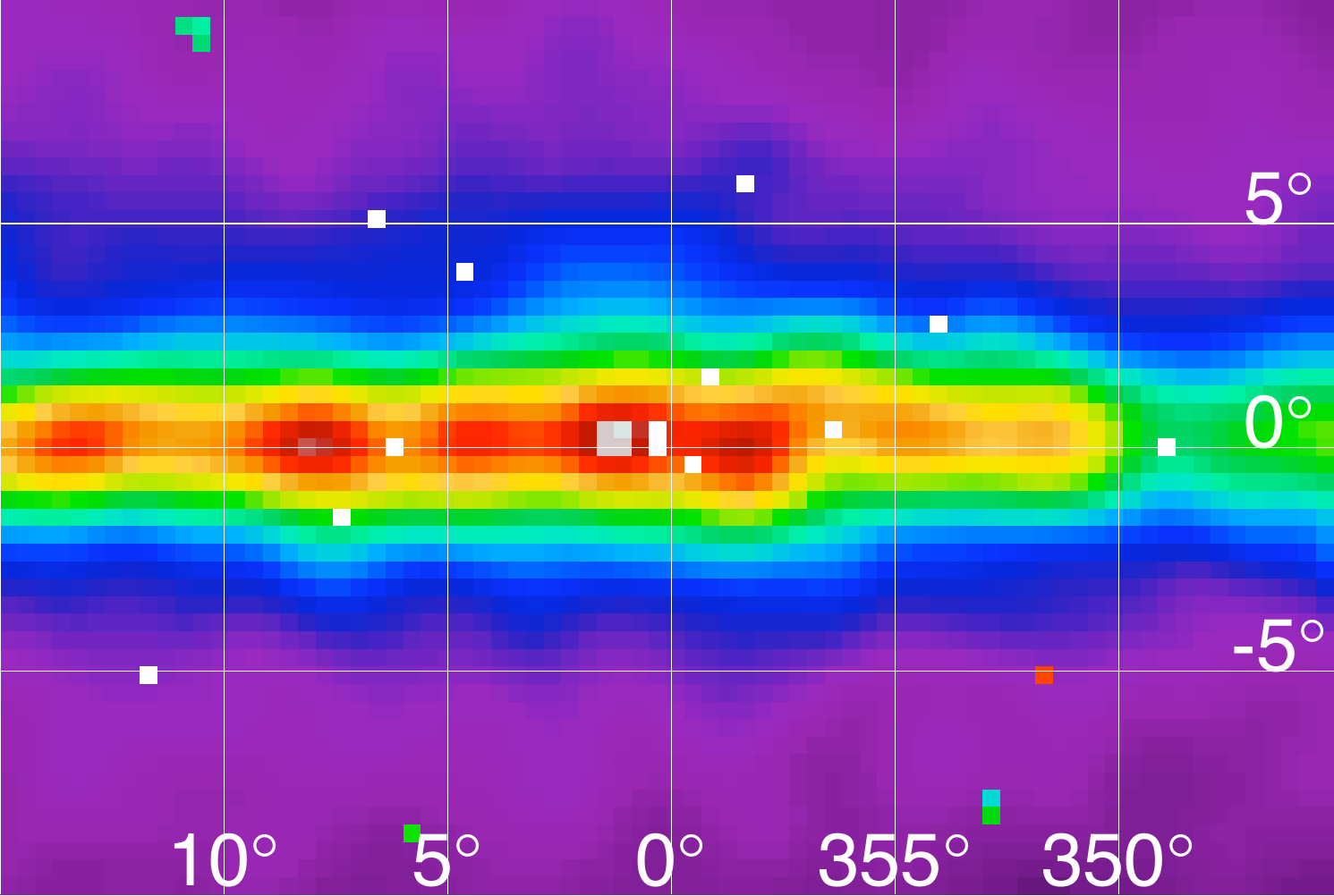}
  \includegraphics[height=3.6cm]{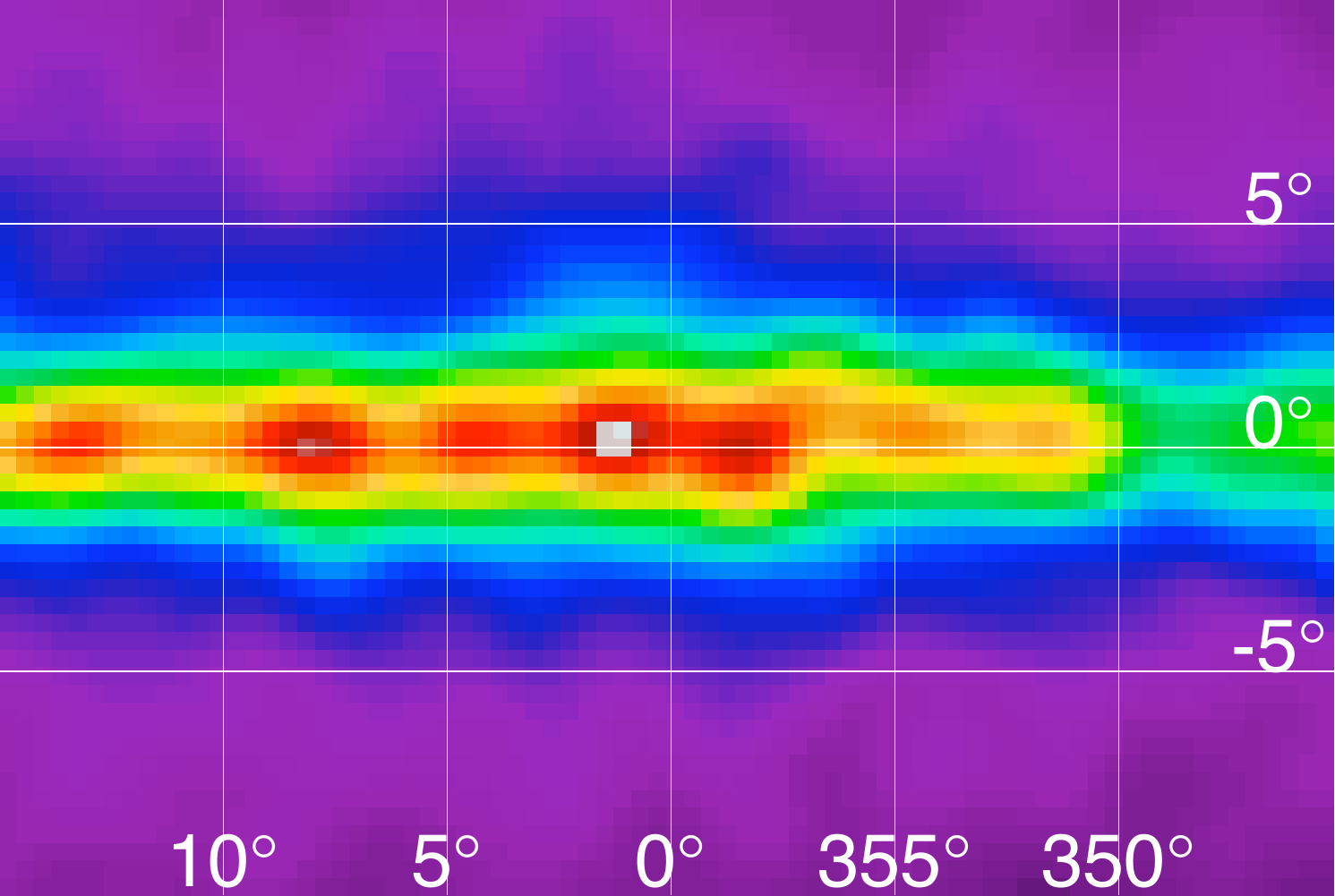}
  \caption{\label{fig:Galactic-plane}Smoothed count map (left), restored image with known point sources included by the dual channel method (middle), and  restored image of the only diffuse channel (right) of the Galactic plane around the Galactic center by EGRET.}
\end{figure}

Although the Milky Way can be resolved by the EGRET due to its large size, smaller scale structure in the Galactic plane is not easy to resolve due to large EGRET PSF and contamination of the point sources.
Our image restoration technique can reduce these effects as shown in Figure~\ref{fig:Galactic-plane}.
The left panel shows the smoothed count map, the middle panel shows the restored image with known point sources included by the dual channel method, and the right panel shows the restored image of only the diffuse channel from the previous image, which is equivalent to the point source removal.
The final EGRET image of the Galactic plane has smooth background and less structures associated with ``point sources'' that may or may not be real point sources.

\section{ Conclusions}
Generalization of a widely-used image restoration technique is introduced to incorporate the rapidly varying PSF in the EGRET and GLAST-LAT although it is not trivial since the wavelet filtering is performed on pixelated images.
We demonstrated the effectiveness of our technique using three fields observed by the EGRET.
Our technique is confirmed to be useful to obtain sharper images of extended sources, to identify source confusions, and to cleanly remove point sources from the image of extended sources.




\end{document}